\documentclass[aps,preprint,amsmath,amssymb]{revtex4}
\usepackage{graphicx}
\usepackage{color}
\def\be{\begin{eqnarray}}
\def\ed{\end{eqnarray}}
\def\e{\epsilon}
\def\te{\tilde{\epsilon}}
\def\ve{\varepsilon}
\def\la{\langle}
\def\ra{\rangle}
\def\non{\nonumber}

\begin{document}

\title{\Large \bf Expectations on $B\to \left(K^*_{0}(1430),K^*_{2}(1430)\right)\phi$ decays}

\author{ \bf \large Chuan-Hung Chen$^{1,2}$\footnote{Email:
physchen@mail.ncku.edu.tw} and Chao-Qiang
Geng$^{3,4}$\footnote{Email: geng@phys.nthu.edu.tw}
 }

\affiliation{ $^{1}$Department of Physics, National Cheng-Kung
University, Tainan 701, Taiwan \\
$^{2}$National Center for Theoretical Sciences, Taiwan\\
$^{3}$Department of Physics, National Tsing-Hua University, Hsinchu
300, Taiwan  \\
$^{4}$Theory Group, TRIUMF, 4004 Wesbrook Mall, Vancouver, B.C V6T
2A3, Canada
 }

\date{\today}

\begin{abstract}
As the annihilation contributions play important roles in solving
the puzzle of the small longitudinal polarizations in $B\to  K^*
\phi$ decays, we examine the similar effects in the decays of $B\to
K^*_{0,2}(1430) \phi$. For the calculations on the annihilated
contributions, we adopt that the form factors in $B\to K^{(*)} \phi$
decays are parameters determined by the observed branching ratios
(BRs), polarization fractions (PFs) and relative angles in
experiments and we connect the parameters between  $B\to K^*_{0(2)}
\phi$ and $B\to K^{(*)}\phi$ by the ansatz of correlating $\la
K^*_n(1430) \phi| (V-A)_{\mu}|0\ra$ to $\la K^{(*)} \phi|
(V-A)_{\mu}|0\ra$.  We find that the BR of $B_d\to K^{*0}_{0}(1430)
\phi$ is $(3.69 \pm 0.47)\times 10^{-6}$. By using the  transition
form factors of $B\to K^*_2(1430)$ in the light-front quark model
(LFQM) and the 2nd version of Isgur-Scora-Grinstein-Wise (ISGW2), we
show that BR of $B_d\to K^{*0}_{2}(1430)\phi$ is a broad allowed
value and $(1.70\pm0.80)\times 10^{-6}$, respectively. In terms of
the recent BABAR's observations on BRs and PFs in $B_d\to
K^{*0}_{2}(1430)\phi $, the results in the LFQM are found to be more
favorable. In addition, due to the annihilation contributions to
$B\to K^*_2\phi$ and $B\to K^*\phi$ being opposite in sign, we
demonstrate that the longitudinal polarization of $B_d\to
K^{*0}_2(1430) \phi$ is always $O(1)$ with or without including the
annihilation contributions.
\end{abstract}

\maketitle %

\section{Introduction}
Since the transverse decay amplitudes of vector meson productions in
$B$ decays are associated with their masses, by naive estimations,
the longitudinal polarization (LP) of $B$ decaying into two light vector
mesons is close to unity. The expectation is confirmed by
BELLE \cite{belle1} and BABAR \cite{babar1-1, babar1-2} in $B\to
\rho(\omega) \rho$ decays, in which the longitudinal parts occupy
over $88\%$. Furthermore, the LPs could be small if the final states
include heavy vector mesons. The conjecture is verified in $B\to
J/\Psi K^*$ decays \cite{belle2,babar2}, in which the longitudinal
contribution is only about $ 60\%$. However, the rule for the small LPs
seems to be violated in $B\to \phi K^*$ decays. From the
measurements of BELLE \cite{belle3} and BABAR
\cite{babar1-1,babar3}, it is quite clear that the LPs in $B\to K^*
\phi$ are only around $50\%$. To solve the unexpected
observations, many
mechanisms have been proposed, such as those with
new QCD effects \cite{Kagan,QCD,Chen,Beneke} as well as
extensions of the standard model (SM)
\cite{newphys,CG_PRD71}.

Recently, the BABAR Collaboration has observed the branching ratios (BRs)
and polarization fractions (PFs) for the decays of
$B_d\to K^{*0}_{0,2}(1430) \phi$ \cite{babar4}, given by
 \be
 BR(B_d\to K^{*0}_{2}(1430) \phi)&=& (7.8\pm 1.1 \pm 0.6)\times 10^{-6}\,, \non\\
 R_{L}(B_d\to K^{*0}_{2}(1430)&=& 0.853 ^{+0.061}_{-0.069}\pm 0.036\,, \non \\
 R_{\perp}(B_d\to K^{*0}_{2}(1430)&=& 0.045^{+0.049}_{-0.040} \pm 0.013\,,
\non\\
BR(B_d\to K^{*0}_{0}(1430) \phi)&=& (4.6\pm 0.7 \pm
0.6)\times 10^{-6}\,.
 \ed
By the observations, it seems that the LP of the
p-wave tensor-meson production  is much larger than
those of the s-wave vector mesons in $B$ decays. To find out
 whether the data are
just the statistical fluctuation or
the correct tendency for the
behavior of the p-wave productions in $B$ decays, it is important
to study the
phenomena from theoretical viewpoint.

It is known that the annihilation contributions play important roles
in the PFs of $B\to K^* \phi$ decays \cite{Kagan,Chen,Beneke}. As
the corresponding time-like form factors are more uncertain than
those of the transition form factors,
we first adopt
that the form factors of the annihilation contributions
on $B\to K^* \phi$ are parameters
 fixed
by the data in $B\to K^* \phi$ and then connect them
to those in $B_d\to K^{*0}_{2}(1430)\phi$. To
have more illustrating examples, we also examine the decays of $B\to
K^*_{0}(1430) \phi$ simultaneously.

The paper is organized as follows. In Sec. II,
 we carry out a general study
on the decay amplitudes and hadronic matrix elements. We present our numerical analysis
in Sec. III. Our conclusions are presented in Sec. IV.

\section{Decay amplitudes and hadronic matrix elements}

It is known that the effective interactions for the decays of $ B\to
K^*_{n}(1430) \phi\ (n=0,2)$ are described by $b\to s q\bar{q}$,
which are the same as $B\to K^{(*)} \phi$,  given by
\cite{BBL}
\begin{equation}
H_{{\rm eff}}={\frac{G_{F}}{\sqrt{2}}}\sum_{q=u,c}V_{q}\left[
C_{1}(\mu) O_{1}^{(q)}(\mu )+C_{2}(\mu )O_{2}^{(q)}(\mu
)+\sum_{i=3}^{10}C_{i}(\mu) O_{i}(\mu )\right] \;,
\label{eq:hamiltonian}
\end{equation}
where $V_{q}=V_{qs}^{*}V_{qb}$ are the Cabibbo-Kobayashi-Maskawa
(CKM)  matrix elements \cite{CKM} and the operators $O_{1}$-$O_{10}$
are defined as
\begin{eqnarray}
&&O_{1}^{(q)}=(\bar{s}_{\alpha}q_{\beta})_{V-A}(\bar{q}_{\beta}b_{\alpha})_{V-A}\;,\;\;\;\;\;
\;\;\;O_{2}^{(q)}=(\bar{s}_{\alpha}q_{\alpha})_{V-A}(\bar{q}_{\beta}b_{\beta})_{V-A}\;,
\nonumber \\
&&O_{3}=(\bar{s}_{\alpha}b_{\alpha})_{V-A}\sum_{q}(\bar{q}_{\beta}q_{\beta})_{V-A}\;,\;\;\;
\;O_{4}=(\bar{s}_{\alpha}b_{\beta})_{V-A}\sum_{q}(\bar{q}_{\beta}q_{\alpha})_{V-A}\;,
\nonumber \\
&&O_{5}=(\bar{s}_{\alpha}b_{\alpha})_{V-A}\sum_{q}(\bar{q}_{\beta}q_{\beta})_{V+A}\;,\;\;\;
\;O_{6}=(\bar{s}_{\alpha}b_{\beta})_{V-A}\sum_{q}(\bar{q}_{\beta}q_{\alpha})_{V+A}\;,
\nonumber \\
&&O_{7}=\frac{3}{2}(\bar{s}_{\alpha}b_{\alpha})_{V-A}\sum_{q}e_{q} (\bar{q}%
_{\beta}q_{\beta})_{V+A}\;,\;\;O_{8}=\frac{3}{2}(\bar{s}_{\alpha}b_{\beta})_{V-A}
\sum_{q}e_{q}(\bar{q}_{\beta}q_{\alpha})_{V+A}\;,  \nonumber \\
&&O_{9}=\frac{3}{2}(\bar{s}_{\alpha}b_{\alpha})_{V-A}\sum_{q}e_{q} (\bar{q}%
_{\beta}q_{\beta})_{V-A}\;,\;\;O_{10}=\frac{3}{2}(\bar{s}_{\alpha}b_{\beta})_{V-A}
\sum_{q}e_{q}(\bar{q}_{\beta}q_{\alpha})_{V-A}\;,
\end{eqnarray}
with $\alpha$ and $\beta$ being the color indices
and $C_{1}$-$C_{10}$ the corresponding Wilson
coefficients (WCs).
In Eq.
(\ref{eq:hamiltonian}), $O_{1}$-$O_{2}$ are from the tree level of
weak interactions, $O_{3}$-$O_{6}$ are the so-called gluon penguin
operators and $O_{7}$-$O_{10}$ are the electroweak penguin
operators. Using the unitarity condition, the CKM matrix
elements for the penguin operators $O_{3}$-$O_{10}$ can also be
expressed by $V_{u}+V_{c}=-V_{t}$. Besides the weak effective
interactions, to study exclusive two-body decays, we should know
how to calculate the transition matrix elements such as $\la M_{1}
M_{2}| H_{\rm eff}|  B\ra$, where nonperturbative effects dominate
the uncertainties. By taking the heavy quark limit, we consider that
the productions of light mesons satisfy the assumption of color
transparency \cite{Bjorken}, i.e., the final state interactions are
subleading effects and negligible. Hence,
the decays of $B\to K^*_{n}(1430)\phi$ could be treated as
short-distance dominant processes. As the wave functions of
p-wave states are quite uncertain, unlike those of s-wave states
which are known  at least in the leading twist-2
and twist-3 \cite{WF_QCD}, in our calculations we will
employ the generalized factorization assumption \cite{GFA1,GFA2}, in
which the factorized parts are regarded as the leading effects and
the nonfactorized effects are lumped and characterized by the
effective number of colors, denoted by $N^{\rm eff}_c$ \cite{BSW}.

\begin{figure}[htbp]
\includegraphics*[width=4.in]{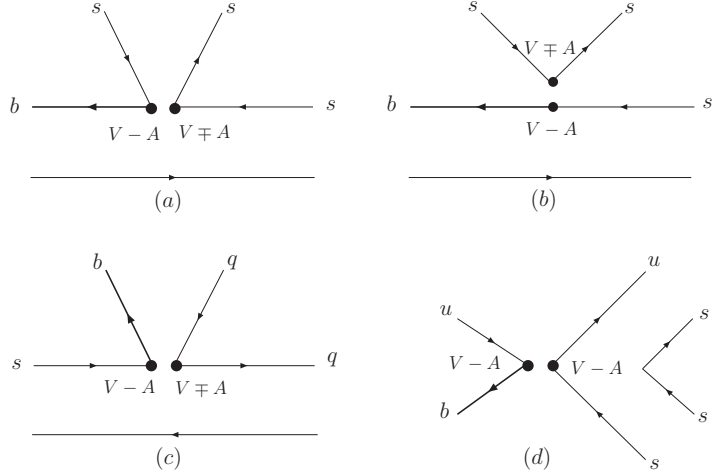}
\caption{Flvaor diagrams for $ B\to \left( K^*_{0}(1430),\,
K^*_{2}(1430)\right)\phi $ decays, where (a) and (b) denote the
penguin emission topologies, while (c) and (d) are the annihilation
topologies for penguin and tree contributions, respectively. }
 \label{fig:flavor-diagram}
\end{figure}

Based on the effective interactions of Eq.~(\ref{eq:hamiltonian}),
the matrix elements $\la K^*_{n}(1430) \phi| H_{\rm eff}| B\ra$
could be classified by various flavor diagrams displayed in
Fig.~\ref{fig:flavor-diagram}, where
(a) and (b) denote the penguin emission topologies, while (c)[(d)]
is the penguin [tree] annihilation topology.
Furthermore, in terms of the flavor diagrams, we can group the
effects of Eq.~(\ref{eq:hamiltonian}) for the transition matrix elements
to be
  \be
    X^{( BK^*_{n},\phi)}&=&\langle \phi|(\bar{s} s)_{V\pm A}|0\rangle \,
   \langle  K^*_n| (\bar{b} s)_{V-A}|B\rangle\,, \nonumber \\
   Z^{(B, K^*_n \phi)}_{1}&=&\langle  K^*_n\phi  |(\bar{q} s)_{V-A}|0\rangle \,
   \langle 0 | (\bar{b} q)_{V-A}|B\rangle\,, \nonumber \\
  Z^{(B, K^*_n \phi)}_{2}&=&\langle  K^*_n\phi |(\bar{q} s)_{S-P}|0\rangle \,
   \langle 0 | (\bar{b} q)_{S+P}|B \rangle\,,
  \label{eq:XZ}
  \ed
where $X^{(BK^*_n,\phi)}$ represent the factorized parts of the
emission
topology and  $Z^{(B, K^*_n \phi)}_{1,2}$ stand for the factorized
parts of the annihilation topology. Note that the currents associated
with $(S+P)\otimes (S-P)$ in Eq. (\ref{eq:XZ}) are from the Fierz
transformations of $(V-A)\otimes (V+A)$. From Eqs.
(\ref{eq:hamiltonian})-(\ref{eq:XZ}), the decay amplitudes for $B\to
K^{*}_{n}(1430) \phi$ can be written as%
\be
  A(B_{d}\to  K^{*0}_{n}(1430) \phi)&=& \frac{G_F}{\sqrt 2}\left\{
  - V_{tb} V^{*}_{ts} \left [
  \tilde{a}^{(s)}
  X^{(BK^{*}_{n},\phi)} \right. \right. \nonumber \\
  && \left. \left.+a^{(s)}_{4} Z^{(B, K^{*}_{n}\phi)}_{1} -2 a^{(s)}_{6}Z^{(B, K^{*}_{n}\phi
 )}_{2}\right ]\right \}, \non
  \\
  A(B^{+}_{u}\to K^{*+}_{n}(1430) \phi)&=& \frac{G_F}{\sqrt 2}\left\{
  V_{us}V_{ub}^{*} a_{1} Z^{(B, K^{*}_{n}\phi)}_{1}- V_{tb} V^{*}_{ts}
  \left[ \tilde{a}^{(s)}
  X^{(BK^{*}_{n},\phi)} \right. \right. \nonumber \\
  && \left. \left.+a^{(u)}_{4} Z^{(B, K^{*}_{n}\phi)}_{1} -2 a^{(u)}_{6} Z^{(B, K^{*}_{n}\phi
 )}_{2}\right ]\right \}\,, \label{eq:amp1}
\ed%
with $\tilde{a}^{(s)}=a^{(s)}_{3}+a^{(s)}_{4}+a^{(s)}_{5}$. To
be more convenient for our analysis, we have redefined the useful
WCs by combining the gluon and electroweak penguin contributions to be
 \be
  a_{1}&=&C^{\rm eff}_{2}+\frac{C^{\rm eff}_1}{N^{\rm eff}_{c}},\ \ \
   a_{2}=C^{ \rm eff}_{1}+\frac{C^{\rm eff}_2}{N^{\rm eff}_{c}}, \ \
    a^{(q)}_{3}=C^{\rm eff}_{3}+\frac{C^{\rm eff}_4}{N^{\rm eff}_{c}}
  +\frac{3}{2}e_{q}\left(C^{\rm eff}_{9}+\frac{C^{\rm eff}_{10}}{N^{\rm eff}_{c}}\right),
        \nonumber \\
  a^{(q)}_{4}&=&C^{\rm eff}_{4}+\frac{C^{\rm eff}_3}{N^{\rm eff}_{c}}+\frac{3}{2}e_{q}\left(C^{\rm eff}_{10}
        +\frac{C^{\rm eff}_{9}}{N^{\rm eff}_{c}}\right),
  a^{(q)}_{5}=C^{\rm eff}_{5}+\frac{C^{\rm eff}_{6}}{N^{\rm eff}_{c}}+\frac{3}{2}e_{q}\left(C^{\rm eff}_{7}
        +\frac{C^{\rm eff}_{8}}{N^{\rm eff}_{c}}\right),\nonumber
        \\
  a^{(q)}_{6}&=&C^{\rm eff}_{6}+\frac{C^{\rm eff}_5}{N^{\rm eff}_{c}}+\frac{3}{2}e_{q}\left(C^{\rm eff}_{8}+\frac{C^{\rm
        eff}_{7}}{N^{\rm eff}_{c}}\right),
  \ed
where the effective WCs of $C^{\rm eff}_{i}$ contain
 vertex corrections
for smearing the $\mu$-scale dependences in the transition matrix
elements \cite{GFA2} and the effective color number of $N^{\rm eff}_{c}$ is
a variable, which may not be equal to $3$
\cite{GFA1,GFA2,BSW}.

The hadronic matrix elements defined in Eq.~(\ref{eq:XZ}) are the
essential quantities that we have to deal with in the two-body
exclusive B decays. In the following discussions, we analyze the
quantities $X^{BK^*_{n},\phi}$, $Z^{B,K^*_n\phi}_{1}$ and
$Z^{B,K^*_n\phi}_{2}$ individually. Since the degrees of freedom of
$K^*_{0}$ are less than those of $K^*_{2}$, we
start with
$B\to K^*_{0}(1430)\phi$. As usual, we define the
various normal hadronic matrix elements as follows:
\cite{LF1}
 \be
\la 0|\bar b \gamma^{\mu} \gamma_{5} q| B(p_B)\ra&=& if_{B}
p^{\mu}_{B}\,, \ \ \ \la \phi(q,\, h)|\bar s \gamma^{\mu} s|0\ra =
m_{\phi} f_{\phi}\ve^{\mu*}_{\phi}(h) \,,
 %
\non\\
 \la  K^*_{0}(p)|V_{\mu} - A_{\mu}|B(p_{B})\ra &=&
 i\left[\left(P_\mu-{m^{2}_{B}-m^{2}_{K^*_0}\over q^2}\,q_
 \mu\right) F_1^{BK^*_{0}}(q^2) \right.\non \\
 &&\left.+{m^{2}_{B}-m^{2}_{K^*_0}\over q^2}
 \,q_\mu\,F_0^{BK^*_{0}}(q^2)\right]\label{eq:ffs}\,,
 %
\ed
where $(V_{\mu },A_{\mu }) =\bar{b}(\gamma _{\mu }, \gamma_{\mu
}\gamma _{5})s$, $m_{B,\phi,K^*_0}$ are the meson masses of $B$,
$\phi$ and $K^*_{0}$, $P=p_{B}+p$, $q=p_{B}-p$ and $P \cdot
q=m^{2}_{B}-m^{2}_{K^*_n}$. Similarly, the time-like
form factors for $\la K^*_{0} \phi| V_{\mu} - A_{\mu}| 0 \ra$ could
be defined by
 \be
 \la  K^*_{0}(p) \phi(q,\ve_{\phi})|V_{\mu} - A_{\mu}|0 \ra &=&
 -i\frac{V^{K^*_0 \phi}(q^{2})}{m_{\phi}-m_{K^*_0}}\e_{\mu \nu \rho
\sigma}\ve_{\phi}^{\nu* }Q^{\rho} s^{\sigma }\,, \non\\
&&+\left\{2m_{\phi}A^{K^*_0 \phi}_{0}(q^{2})\frac{\ve_{\phi}
^{*}\cdot Q}{ Q^{2}}Q_{\mu
} \right.\non \\
&&+( m_{\phi} -m_{K^*_0}) A^{K^*_0 \phi}_{1}(q^{2})\Big(\ve_{\phi\mu
}^{*}
-\frac{\ve_{\phi}^{*}\cdot Q}{Q^{2}}Q_{\mu }\Big)\nonumber\\
&&\left.-A^{K^*_0 \phi}_{2}(q^{2})\frac{\ve_{\phi}^{*}\cdot
Q}{m_{\phi}-m_{K^*_0}}
\Big(s_{\mu } - \frac{s\cdot Q}{Q^{2}}Q_{\mu } %
\Big)\right\} \label{eq:vs}
 \ed
 with $Q_{\mu}=(p+q)_{\mu}=p_{B\mu}$ and $s_{\mu}=(p-q)_{\mu}$.
In terms of form factors in Eqs. (\ref{eq:ffs}) and (\ref{eq:vs}),
Eq.~(\ref{eq:XZ}) could be rewritten as
\be
    X^{( BK^*_{0},\phi)}&=&i2m_{\phi} f_{\phi} F^{BK^*_0}_{1}(m^2_{\phi})
    \ve^{*}_{\phi} \cdot p_{B}\,, \non \\
   Z^{(B, K^*_0 \phi)}_{1}&=&-i2m_{\phi} f_{B} A^{K^*_0 \phi}_{0}(m^2_{B})
   \ve^{*}_{\phi}\cdot p_{B}\,, \non \\
  Z^{(B, K^*_0 \phi)}_{2}&=&i 2m_{\phi} f_{B}\frac{m^2_{B}}{(m_{b}+m_{q})(m_s-m_q)}
  A^{K^*_0 \phi}_{0}(m^2_{B})\ve^{*}_{\phi}\cdot p_{B} \,.
  \label{eq:XZ2}
  \ed
To compare with the s-wave states, we also give
the hadronic matrix elements in $B\to K\phi$
as
 \be
    X^{( BK,\phi)}&=&2m_{\phi} f_{\phi} F^{BK}_{1}(m^2_{\phi})
    \ve^{*}_{\phi} \cdot p_{B}\,, \non \\
   Z^{(B, K \phi)}_{1}&=&-2m_{\phi} f_{B} A^{K \phi}_{0}(m^2_{B})
   \ve^{*}_{\phi}\cdot p_{B}\,, \non \\
  Z^{(B, K \phi)}_{2}&=&- 2m_{\phi} f_{B}\frac{m^2_{B}}{(m_{b}+m_{q})(m_s+m_q)}
  A^{K \phi}_{0}(m^2_{B})\ve^{*}_{\phi}\cdot p_{B} \,.
  \label{eq:XZ_Kphi}
  \ed
We note that except the factor of $-i$
associated with the p-wave states \cite{LF1}, the definitions of
the form factors for $\la K| V_{\mu}-A_{\mu}|B \ra$ and $\la K \phi|V_{\mu} -
A_{\mu}|0\ra$ are similar to those for $\la K^*_0| V_{\mu}-A_{\mu}|
B\ra$ and $\la K^*_0 \phi|V_{\mu} - A_{\mu}|0\ra$, respectively.
 We will discuss
the behaviors of $A^{K^*_0\phi}_0$ and $A^{K\phi}_{0}$ later on.

We now investigate the decays of $B\to K^*_2(1430) \phi$,
which are similar to $B\to K^* \phi$.
The analogy of Eq.~(\ref{eq:XZ}) for $B\to K^* \phi$ can be presented by
   \be
    X^{( BK^*,\phi)}&=& -i m_{\phi} f_{\phi} \left\{
    (m_{B}+m_{K^*})A^{BK^*}_{1}(q^2) \ve^*_{\phi}\cdot \ve^*_{K^*}
    - \frac{2A^{BK^*}_{2}(q^2)}{m_{B}+m_{K^*}}\ve^*_{\phi}\cdot p_{B} \ve^*_{K^*}\cdot
    p_B \right. \non\\
    &&\left.+i \frac{2V^{BK^*}(q^2)}{m_{B}+m_{K^*}} \e_{\mu \nu \rho \sigma}
\ve^{\mu*}_{\phi} \ve^{\nu*}_{K^*} q^{\rho} p^{\sigma}
    \right\}\,, \non \\
   Z^{(B, K^* \phi)}_{1}&=&-if_B\left\{ m^2_{B} V^{K^*\phi}_{1}(Q^2) \ve^*_{\phi}\cdot \ve^*_{T}
   -V^{K^*\phi}_{2}(Q^2) \ve^*_{\phi}\cdot Q \ve^*_{K^*}\cdot
   \right. Q\non\\
   &&\left.+i2A^{K^*\phi}(Q^2)\e_{\mu \nu \rho \sigma} \ve^{\mu*}_{\phi}
   \ve^{\nu*}_{T} q^{\rho} p^{\sigma} \right\}\,, \non \\
  Z^{(B, K^* \phi)}_{2}&=& -i\frac{m^2_{B}f_{B}}{m_{b}+m_{q}} \left\{
    \frac{m^2_{B} V^{K^*\phi}_{1}(Q^2)}{m_s-m_q}\ve^*_{\phi}\cdot \ve^*_{K^*}
  -\frac{V^{K^*\phi}_{2}(Q^2)}{m_s-m_q}\ve^*_{\phi}\cdot Q \ve^*_{K^*}\cdot Q
  \right.\non\\
  &&\left.+i2\frac{A^{K^*\phi}(Q^2)}{m_s+m_q}\e_{\mu \nu \rho \sigma} \ve^{\mu*}_{\phi}
   \ve^{\nu*}_{K^*} q^{\rho} p^{\sigma}\right\}\,,
  \label{eq:XZV}
  \ed
where we have used the form factors in the transition
of $B\to K^*$ and $\la K^*
\phi| V_{\mu}(A_{\mu}) |0\ra$, defined by \cite{LF1}
 \be
\la  K^*(p,\ve_{K^*} )| V_{\mu }| B%
(p_{B})\rangle &=&-\frac{V^{BK^*}(q^{2})}{m_{B}+m_{K^*}}\e_{\mu \nu
\rho \sigma}\ve_{K^*} ^{\nu* }P^{\rho }q^{\sigma },
\nonumber \\
\la K^*(p,\ve_{K^*} )| A_{\mu }| B(p_{B})\rangle
&=&i\left\{2m_{V}A^{BK^*}_{0}(q^{2})\frac{\ve_{K^*} ^{*}\cdot q}{
q^{2}}q_{\mu
} \right.\non \\
&&+(m_{B}+m_{V}) A^{BK^*}_{1}(q^{2})\Big(\ve_{V\mu }^{*}
-\frac{\ve_{K^*}^{*}\cdot q}{q^{2}}q_{\mu }\Big)\nonumber\\
&&\left.-A^{BK^*}_{2}(q^{2})\frac{\ve_{K^*} ^{*}\cdot q}{m_{B}+m_{K^*}}\Big( P_{\mu }-%
\frac{P\cdot q}{q^{2}}q_{\mu }\Big)\right\}\, ,
  \label{eq:vff}
 \ed
and \cite{EDS}
 \be %
 \la K^* \phi| \bar q \gamma_{\mu}\gamma_5 s Q^{\mu}
 | 0 \ra &=& -iA^{K^*
\phi}(q^2) \e_{\mu \nu \rho \sigma}\ve^{\mu*}_{\phi}
\ve^{\nu*}_{K^*} Q^{\rho} s^{\sigma}\,, \non\\
 \la K^* \phi| \bar q \gamma_{\mu} s Q^{\mu}|0 \ra &=& m^2_{B} V^{K^*
 \phi}_{1}(q^2) \ve^{*}_{\phi}\cdot \ve^{*}_{K^*} - V^{K^*\phi}_{2}(q^2) \ve^*_{\phi} \cdot Q
 \ve^{*}_{K^*}\cdot Q\,,
 \ed%
respectively. Here, $\ve_{K^*}$ denotes the polarization vector of
the $K^*$ meson.
 To study the production of a tensor meson such as
$K^*_{2}(1430)$ in $B$ decays, we need to introduce the properties
of polarization vectors for the tensor meson.
It is known that the
polarization tensor $\te^{\mu\nu}$ of a tensor meson satisfies
\begin{eqnarray}
\te^{\mu\nu}(p,h)&=& \te^{\nu\mu}(p,h)\,, \ \ \
\te^{\mu\nu}(p,h)p_{\nu}=\te^{\mu\nu}(p,h)p_{\mu}=0\,, \ \ \ g_{\mu
\nu}\te^{\mu\nu}=0\,,
\end{eqnarray}
where $h$ is the meson helicity.
The states of a massive spin-2 particle can be constructed by using
two spin-1 states.
To analyze PFs in the production of the tensor mesonic $B$ decays,
we explicitly express $\te^{\mu\nu}(p,h)$ to be \cite{BDDN}
\begin{eqnarray}
\te^{\mu\nu}(\pm 2)&=&e^{\mu}(\pm) e^{\nu}(\pm) \,,\nonumber \\
\te^{\mu\nu}(\pm 1)&=&\frac{1}{\sqrt{2}}\left[ e^{\mu}(\pm) e^{\nu}(0)+e^{\mu}(0) e^{\nu}(\pm)  \right]\,,\nonumber \\
\te^{\mu\nu}(0)&=& \frac{1}{\sqrt{6}}\left[ e^{\mu}(+)e^{\nu}(-) +
e^{\mu}(-)e^{\nu}(+)\right]+\sqrt{\frac{2}{3}}e^{\mu}(0) e^{\nu}(0)
\,,
\end{eqnarray}
where $e^{\mu}(0,\pm)$ denote the polarization vectors of a massive
vector state and their representations are chosen as
\begin{eqnarray}
e^{\mu}(0)=\frac{1}{m_{T}}(|\vec{p}|,0, 0, E_{T})\,,\ \ \
e^{\mu}(\pm)=\frac{1}{\sqrt{2}}(0, \mp 1, -i, 0)
\end{eqnarray}
with $m_{T}(|\vec{p}|)$ being the mass (momentum) of the particle.
Since the $B$ meson is a spinless particle, the helicities carried by decaying particles in the two-body $B$ decay should be the same.
Moreover, although the tensor meson contains 2 degrees of freedom, only $h=0$ and $\pm 1$ give the contributions. Hence,
it will be useful to redefine the new polarization vector of a
tensor meson to be $\ve_{T\mu}(h)\equiv \te_{\mu\nu}(p,h)
p^{\nu}_{B}$, where
 \be \ve_{T}^{\mu}(\pm 2)=0\,, \ \ \ \ve_{T}^{\mu}(\pm
1)=\frac{1}{\sqrt{2}}e(0)\cdot p_{B} e^{\mu}(\pm)\,,\ \ \
\ve_{T}^{\mu}( 0)=\sqrt{\frac{2}{3}}e(0)\cdot p_{B} e^{\mu}(0)\,,
 \ed
with $e(0)\cdot p_{B}=m_{B}|\vec{p}|/m_{T}$. Based on the new
polarization vector $\ve^{*}_{T}$, the transition form factors for
$B\to K^*_{2}$ could be defined by \cite{LF1}
 \be
\langle  K^*_{2}(p,\ve_{T})| V_{\mu} | B(p_B)\rangle&=&h(q^2)
\e_{\mu\nu\rho
\sigma}\ve^{\nu*}_{T} P^{\rho}q^{\sigma}\,,\nonumber \\
\langle  K^*_{2}(p,\ve_{T})| A_{\mu} | B(p_B)\rangle&=&-i\left[
k(q^2)\ve^*_{T\mu}- \ve^*_{T}\cdot P \left(  P_{\mu} b_{+}(q^2) -
q_{\mu} b_{-}(q^2)\right)
\right] \,,
%
 \ed
and  the time-like form factors for $\la K^*_{2} \phi| V_{\mu} -
A_{\mu}| 0 \ra$ could be parametrized as
 \be
\la  K^*_{2}(p,\ve_T) \phi(q,\ve_{\phi})|(V_{\mu} - A_{\mu})
Q^{\mu}|0 \ra &=& -iA^{K^*_2 \phi}(Q^{2})\e_{\mu \nu \rho
\sigma}\ve_{\phi}^{\mu* } \ve^{\nu*}_{T}Q^{\rho} s^{\sigma } \non\\
&&- V^{K^*_2 \phi}_{1}(Q^2) \ve^*_{\phi}\cdot \ve^*_T+ V^{K^*_2
\phi}_{2}(Q^2) \ve^*_{\phi}\cdot Q \ve^*_{T}\cdot Q \,.
\label{eq:vT}
 \ed
Consequently, the analogy of
Eq.~(\ref{eq:XZ}) for $B\to K^*_2(1430) \phi$ could be
explicitly expressed by%
\be
    X^{( BK^*_2,\phi)}&=& i m_{\phi} f_{\phi} \left\{
    k(q^2) \ve^*_{\phi}\cdot \ve^*_{T} - 2b_{+}(q^2) \ve^*_{\phi}\cdot p_B \ve^*_{T}\cdot
    p_B +i 2h(q^2) \e_{\mu \nu \rho \sigma} \ve^{\mu*}_{\phi} \ve^{\nu*}_{T} q^{\rho} p^{\sigma}
    \right\}\,, \non \\
   Z^{(B, K^*_2 \phi)}_{1}&=&if_B\left\{ V^{K^*_2\phi}_{1}(Q^2) \ve^*_{\phi}\cdot \ve^*_{T}
   -V^{K^*_2\phi}_{2}(Q^2) \ve^*_{\phi}\cdot Q \ve^*_{T}\cdot
   \right. Q\non\\
   &&\left.+i2A^{K^*_2\phi}(Q^2)\e_{\mu \nu \rho \sigma} \ve^{\mu*}_{\phi}
   \ve^{\nu*}_{T} q^{\rho} p^{\sigma} \right\}\,, \non \\
  Z^{(B, K^*_2 \phi)}_{2}&=& -i\frac{m^2_{B}f_{B}}{m_{b}+m_{q}} \left\{
    \frac{V^{K^*_2\phi}_{1}(Q^2)}{m_s-m_q}\ve^*_{\phi}\cdot \ve^*_{T}
  -\frac{V^{K^*_2\phi}_{2}(Q^2)}{m_s-m_q}\ve^*_{\phi}\cdot Q \ve^*_{T}\cdot Q
  \right.\non\\
  &&\left.+i2\frac{A^{K^*_2\phi}(Q^2)}{m_s+m_q}\e_{\mu \nu \rho \sigma} \ve^{\mu*}_{\phi}
   \ve^{\nu*}_{T} q^{\rho} p^{\sigma}\right\}\,.
  \label{eq:XZT}
  \ed

Since the final sates in the decays $B\to VV$ and $B\to TV$ carry
spin degrees of freedom, the decay amplitudes in terms of helicities
can be generally described by \cite{BVV}
\begin{eqnarray*}
{\cal M}^{(h)} &=&\ve_{1\mu}^{*}(h)\ve_{2\nu}^{*}(h) \left[ a \,
g^{\mu\nu} +\frac{b}{m_{1} m_{2}}\;  p_B^\mu p_B^\nu + i\,
\frac{c}{m_1 m_2}\; \epsilon^{\mu\nu\alpha\beta} p_{1\alpha}
p_{2\beta}\right]\;, \label{eq:hamp}
\end{eqnarray*}
which
can be
decomposed in terms of
\begin{eqnarray}
H_{00}&=&- \left[ax+b\left(x^2-1 \right) \right]\,, \nonumber\\
H_{\pm\pm}&=& a\pm c \sqrt{x^2-1})\,, \label{eq:helicity_Ksphi}
\end{eqnarray}
and
\begin{eqnarray}
H_{00}&=&- \sqrt{\frac{2}{3}} e(0)\cdot p_{B} \left[ax+b\left(x^2-1 \right) \right], \nonumber\\
H_{\pm\pm}&=& \frac{1}{\sqrt{2}} e(0)\cdot p_{B} \left( a\pm c
\sqrt{x^2-1}\right) \label{eq:helicity_K2phi}
\end{eqnarray}
for $B\to K^* \phi$ and $B\to K^*_2 \phi$, respectively, with
$x=(m^2_B-m^2_1-m^2_2)/(2m_1 m_2)$. In addition, we can also write
the amplitudes in terms of polarizations as
\begin{eqnarray}
A_{L}=H_{00} \ \ \ A_{\parallel(\perp)}=\frac{1}{\sqrt{2}}(H_{++}
\pm H_{--}). \label{eq:pol-amp}
\end{eqnarray}
As a result, the BRs  are given by
 \be
  BR(B\to M\phi)=\frac{|\vec{p}|}{8\pi m^2_{B}}
   \left( |A_L|^2+|A_{\parallel}|^2+|A_{\perp}^2|\right)
 \ed
where $M=(K^*,\, K^*_2(1430))$ and $|\vec{p}|$ is the magnitude of
the outgoing momentum, and the corresponding
PFs can be defined to be %
 \be
  R_i&=&\frac{|A_i|^2}{|A_L|^2+|A_{\parallel}|^2+|A_{\perp}^2|}\,,\ \
(i=L,\parallel,\perp)\,, \label{eq:pol}
 \ed %
representing longitudinal, transverse parallel and transverse
perpendicular components, respectively. Note that $\sum_i R_i=1$.

\section{Numerical analysis}
In Tables \ref{tab:decay_con}
and \ref{tab:ff},
we display the meson decay constants and
the transition form factors, respectively.
 Since the numerical values of $B\to K^*_2$ in the LFQM are
different from those in the 2nd version of
the Isgur-Scora-Grinstein-Wise approach (ISGW2) \cite{ISGW2,KLO_PRD67},
in the Table~\ref{tab:ff} we list
 both results.
\begin{table}[hptb]
\caption{Meson decay constants (in units of GeV) of
meson.}\label{tab:decay_con}
\begin{ruledtabular}
\begin{tabular}{cccccc}
$f_{K}$ & $f_{B}$ & $f_{K^*}$ & $f^{T}_{K^*}$ & $f_{\phi}$ &
$f^{T}_{\phi}$
 \\ \hline
$0.16$ & $0.19$ & $0.20$ & $0.16$ & $0.23$ & $0.20$ \\
\end{tabular}
\end{ruledtabular}
\end{table}
\begin{table}[hptb]
\caption{ Transition form factors by the LFQM and
ISGW2.}\label{tab:ff}
\begin{ruledtabular}
\begin{tabular}{ccccccccc}
$F(m^2_{\phi})$ & $F^{BK}_{1}$ &  $V^{BK^*}$ & $A^{BK^*}_{1}$ & $A^{BK^*}_{2}$ &
$F^{BK^*_0}_{1}$ & $k$ & $b_{+}$ & $h$
 \\ \hline
 LFQM & $0.37$ & $0.33$ & $0.27$ & $0.26$ & $0.275$ & $0.013$ & $0.0065$& $0.0087$
\\ \hline
ISGW2 & & & & & & 0.217 & 0.0045 & 0.0045 \\
\end{tabular}
\end{ruledtabular}
\end{table}
In Table~\ref{tab:brpf_noani}, we show
the results without the
annihilated topologies  with
various effective $N^{\rm eff}_{c}$, where
 the $\mu$ scale for the effective Wilson's
coefficients is fixed to be $\mu=2.5$ GeV which is usually adopted
in the literature.
For an explicit example of the effective WCs, we have that
$\tilde{a}^{(s)}(\mu=2.5\, \rm GeV)=(-584-97i,\,
-418-73i,\, -284-55i,\, 84-27i)\times 10^{-4}$ for $N^{\rm eff}_c=
2, 3, 5 ,\infty$, respectively.
 In our naive estimations, we see
that the BRs for $B_{d}\to ( K^{(*)0},\, K^{*0}_{n}(1430)) \phi$
decays are close to the world average values when $N^{\rm
eff}_{c}=3$. This could indicate that the nonfactorized
contributions in the
 processes are small. However, from
Table~\ref{tab:brpf_noani}, the longitudinal and transverse PFs for
$B_{d}\to  K^{*0} \phi$ are inconsistent with the measurements. In
addition, we find that
$R_{\parallel}=1-R_{L}-R_{\perp}$ of  $B_{d}\to
K^{*0}_{2}(1430)\phi$ in the LFQM  almost vanishes
due to $k(m^2_{\phi})_{LFQM}<<k(m^2_{\phi})_{ISGW2}$.
\begin{table}[hptb]
\caption{ BRs (in units of $10^{-6}$) and PFs without annihilation
contributions in the LFQM [ISGW2].}\label{tab:brpf_noani}
\begin{ruledtabular}
\begin{tabular}{ccccccc}
 Mode & (BR, PF) &  $N^{\rm eff}_{c}=2$ & $N^{\rm eff}_{c}=3$ &
 $N^{\rm eff}_{c}=5$ & $N^{\rm eff}_{c}=\infty$ & Exp.
 \\ \hline
 $B_{d}\to \bar K^{0} \phi$ & BR & $15.77$ & $8.10$ & $3.77$ & $0.35$ & $8.3^{+1.2}_{-1.0}$  \\
  \hline
 $B_{d}\to \bar K^{*0}_{0} \phi$ &$BR$ & $7.70$ & $3.95$ & $1.84$ & $0.17$ & $4.6\pm 0.7 \pm0.6$ \\
  \hline
 $B_d\to \bar K^{*0}\phi$ & $BR$ & $14.78$ & 7.59 & 3.52 & 0.33 & $9.5\pm 0.9$\\
  &   $R_L$ & 0.90 & 0.90 & 0.90 & 0.90 & $0.49 \pm 0.04$\\
  &  $R_{\perp}$ & 0.04 & 0.04 & 0.04 & 0.04 & $0.27^{+0.04}_{-0.03}$
 \\ \hline
 $B_d\to \bar K^{*0}_{2}\phi$ & $BR$ & $13.64 [7.77]$ & 7.0 [3.99] & 3.26 [1.86] & 0.30[0.17]& $7.8\pm 1.1 \pm 0.6$ \\
  &   $R_L$ & 0.96 [0.90] & 0.96 [0.90] & 0.96 [0.90] & 0.96 [0.90] & $0.853^{+0.061}_{-0.069}\pm 0.036$\\
  &  $R_{\perp}$ & 0.04 [0.02] & 0.04 [0.02] & 0.04 [0.02] & 0.04
  [0.02] & $0.045^{+0.049}_{-0.040} \pm 0.013$
 \\
\end{tabular}
\end{ruledtabular}
\end{table}

It has been concluded that the annihilation contributions could
significantly reduce the longitudinal polarizations in $B\to K^*
\phi$ decays \cite{Kagan,Chen,Beneke}. It is interesting to ask
whether such effects could also play important roles on
BRs and PFs in $B\to K^*_n(1430) \phi$.
To answer the question, we start with the annihilation contributions
in $B\to PP$, which could provide useful information on the general
properties of the time-like form factors.
In the decays, the
factorized amplitude associated with the $(V-A)\otimes (V-A)$
interaction for the annihilated topology is given by \cite{CGHW}
\begin{eqnarray}
\langle P_{1} P_{2} | \bar{q}_{1}\gamma^{\mu}(1-\gamma_5) q_{2}\,
\bar{q}_{3} \gamma^{\mu} (1-\gamma_5) b | B \rangle_{a}=-if_{B}
(m^{2}_{1}-m^{2}_{2})F^{P_1 P_2}_{0}(m^{2}_{B})\,,
\label{eq:anni1}
\end{eqnarray}
where $m_{1,2}$ are the masses of outgoing particles and $F^{P_1
P_2}_{0}(m^2_{B})$ correspond to the time-like form factor, defined
by
 \begin{eqnarray}
  \langle P_{1}(p_1) P_{2}(p_2) | \bar{q}_{1} \gamma_{\mu} q_{2} |
 0\rangle & =& \left[q_{\mu}-\frac{m^{2}_1 -m^{2}_{2}}{Q^2}Q_{\mu}
 \right] F^{P_1 P_2}_{1}(Q^2) + \frac{m^2_1-m^2_2}{Q^2}Q_{\mu} F^{P_1
 P_2}_{0}(Q^2)
\end{eqnarray}
with $q=p_1-p_2$ and $Q=p_1+p_2$. From Eq. (\ref{eq:anni1}), it is
clear that if $m_{1}= m_{2}$, the factorized effects of the annihilation
topology vanish. Consequently, it is concluded that the annihilated
effects associated with $(V-A)\otimes (V-A)$ are suppressed and
negligible. The conclusion could be extended to any process in
two-body $B$ decays \cite{CGHW}. Hence, in the following discussions
we will neglect the contributions of $Z^{B,M\phi}_{1}$
($M=K^{(*)},K^*_n$). However, if the associated interactions are
$(S+P)\otimes (S-P)$, by equation of motion, the decay amplitude
becomes
\begin{eqnarray}
\langle P_{1} P_{2} | \bar{q}_{1}(1+\gamma_5) q_{2}\, \bar{b}
(1-\gamma_5) q_3 | \bar{B}
\rangle_{a}=-if_{B}\frac{(m^{2}_{1}-m^{2}_{2}) m^{2}_{B}}{(m_{q_1} -
m_{q_2})(m_{b}+m_{q_3})} F^{P_1 P_2}_{0}(m^{2}_{B}).
\end{eqnarray}
We see that the subtracted factors appear in the numerator and
denominator simultaneously. As a result, the annihilation effects by
$(S+P)\otimes (S-P)$ interactions can be sizable due to the
cancelation smeared by $(m^2_{1}-m^{2}_{2})/(m_{q_1}-m_{q_2})
\propto (m_{1}+m_{2})$. In comparing with the emission topologies, now the
annihilations associated with $(S-P)\otimes (S+P)$ are only
suppressed by the factor of $m_{1,2}/m_{B}$.

Due to the tree contributions arising from the annihilation, as
analyzed before, their effects could be neglected. Moreover, except
the
 lifetime, there are no differences in the decay
amplitudes between charged and neutral B mesons. Thus, in our
numerical estimations, we just concentrate on the neutral modes.
According to Eqs.~(\ref{eq:amp1}), (\ref{eq:XZ2}) and
(\ref{eq:XZ_Kphi}), the decay amplitudes for $B_d\to K^{*0}_0(1430)
\phi$ and $B_d\to K^{0}\phi$ are given by
 \be
 A(B_d\to
K^{*0}_0(1430) \phi)&=& \frac{G_F}{\sqrt{2}} \tilde{a}^{(s)}
2m_{\phi} f_{\phi} F^{BK^*_{0}}(m^2_{\phi}) \ve^*_{\phi}\cdot p_B
\left[ 1-2r_{a} \frac{A^{K^*_0
\phi}_{0}(m^2_{B})}{F^{BK^*_0}_{1}(m^2_{\phi})}\right]\,,\non \\
  A(B_d\to
K^{0} \phi)&=& \frac{G_F}{\sqrt{2}} \tilde{a}^{(s)} 2m_{\phi}
f_{\phi} F^{BK} (m^2_{\phi}) \ve^*_{\phi}\cdot p_B \left[ 1+2r_{a}
\frac{A^{K
\phi}_{0}(m^2_{B})}{F^{BK}_{1}(m^2_{\phi})}\right]\,,
\label{eq:ampf_Kphi}
 \ed
respectively, where $r_{a}=(a^{(s)}_{6}/\tilde{a}^{(s)}) m^2_{B} f_{B}/(m_b m_sf_{\phi})$. Similarly, in terms of the
helicity basis and
Eq.~(\ref{eq:hamp}), the corresponding amplitudes for $a$, $b$ and
$c$ are given by
 \be
 a^{K^* \phi}&=&\frac{G_F}{\sqrt{2}}V_{tb} V^{*}_{ts}
 \tilde{a}^{(s)} m_{\phi} f_{\phi}
 (m_B+m_{K^*})A^{BK^*}_{1}(m^2_{\phi})  \left[ 1+2R_a \frac{m_{B}V^{K^*\phi}_{1}(m^2_{B})}{(m_B+m_{K^*})A^{BK^*}_{1}(m^2_{\phi})}\right]\,,
 \non \\
 b^{K^* \phi}&=& -\frac{G_F}{\sqrt{2}}V_{tb} V^{*}_{ts}
 \tilde{a}^{(s)} m^2_{\phi} m_{K^*} f_{\phi} \frac{2A^{BK^*}_{2}(m^2_{\phi})}{m_B+m_K^*}\left[ 1+R_{a}
 \frac{(m_B+m_{K^*})V^{K^*\phi}_{2}(m^2_{B})}{m_BA^{BK^*}_{2}(m^2_{\phi})}\right]\,,\non \\
 c^{K^*\phi}&=& \frac{G_F}{\sqrt{2}}V_{tb} V^{*}_{ts}
 \tilde{a}^{(s)} m^2_{\phi} m_{K^*} f_{\phi}
 \frac{2V^{BK^*}(m^2_{\phi})}{ m_B+m_{K^*}}  \left[ 1+2R_{a}\frac{(m_B+m_{K^*})A^{K^*\phi}(m^2_{B})}{m_B V^{BK^*}(m^2_{\phi})}\right]\,,
 \label{eq:ampf_Ksphi}
 \ed
for $B_d\to  K^{*0} \phi$ and
 \be
 a^{K^*_2 \phi}&=&\frac{G_F}{\sqrt{2}}V_{tb} V^{*}_{ts}
 \tilde{a}^{(s)} m_{\phi} f_{\phi}
 k(m^2_{\phi})
 \left[ 1+2R_a \frac{V^{K^*_2\phi}_{1}(m^2_{B})}{m_B k(m^2_{\phi})}\right]\,,
 \non \\
 b^{K^*_2 \phi}&=& -\frac{G_F}{\sqrt{2}}V_{tb} V^{*}_{ts}
 \tilde{a}^{(s)} 2m^2_{\phi} m_{K^*_2} f_{\phi} b_{+}(m^2_{\phi}) \left[ 1+R_{a}
 \frac{V^{K^*_2\phi}_{2}(m^2_{B})}{m_B b_{+}(m^2_{\phi})}\right]\,,\non \\
 c^{K^*_2\phi}&=& \frac{G_F}{\sqrt{2}}V_{tb} V^{*}_{ts}
 \tilde{a}^{(s)} 2m^2_{\phi} m_{K^*_2} f_{\phi}
 h(m^2_{\phi})
  \left[ 1+2R_{a}\frac{A^{K^*_{2}\phi}(m^2_{B})}{m_B h(m^2_{\phi})}\right]\,,
  \label{eq:ampf_K2phi}
 \ed
for $B_d\to  K^*_{2}(1430) \phi$, respectively, where
 \begin{equation}
  R_{a}=\frac{a^{(s)}_{6}}{\tilde{a}^{(s)}} \frac{
m^3_{B}f_{B}}{m_{\phi} m_b m_sf_{\phi}}\,.
 \end{equation}
We note that although the formulas for $K^{{*}0}\phi$ and
$K^*_{2}(1430) \phi$ are the same,  the signs for the time-like
form factors could be different.

To calculate exclusive B decays,  we face the
theoretical uncertainties, such as those from CKM matrix elements, decay constants
and transition form factors. However, these
uncertainties
could be fixed by the experimental data such as those in
the semileptonic decays. In our concerned processes, in fact, the
challenge one is how to get the proper information on
the  time-like form
factors for the annihilation contributions. Since the values of
the time-like form factors are taken at $q^2=m^2_{B}$, in
principle, we can employ the perturbative QCD (PQCD) to do the
calculations \cite{PQCD1,PQCD2}. Unfortunately, it is known
that the predictions of the PQCD on  $R_L$ of $B\to K^* \phi$ are
much larger than the measured values in the data \cite{CKLPRD66,Chen}. As
there exist no better methods to evaluate the time-like form
factors at the moment, in our approach we regard them in $B\to
K^{(*)} \phi$ as free parameters
and  we  determine their
allowed ranges by utilizing the experimental data, such as BRs,
$R_L$ and $R_{\perp}$ for $B\to K^{(*)} \phi$.
However, in general,
 since the time-like
form factors are complex, while there are
 only four observables can be used so far,
we have to reduce the free parameters. It is known that since the
weak phase is very small in $b\to sq\bar {q}$ processes, CP
asymmetries in the SM should be negligible. On the other hand, as we
will not consider CP violation in the processes, the encountering
problems could be simplified by setting the time-like form factors
to be real. In addition, our simplification is supported by the
analysis of Ref.~\cite{SCET} in which at the lowest order in
$\alpha_{s}$ the annihilation amplitudes are real. Once the
parameters in $B\to K^* \phi$ are fixed, we can adopt the ansatz for
the corresponding quantities in $B\to K^*_n(1430) \phi$ to be
 \be
 \frac{|A^{K
\phi}(m^2_{B})|}{F^{BK}_{1}(m^2_{\phi})}&\approx & \frac{|A^{K^*_0
\phi}(m^2_{B})|}{F^{BK^*_0}_{1}(m^2_{\phi})}\,, \non\\
 \frac{m^2_{B} |V^{K^*\phi}_{1}(m^2_{B})|}{(m_B+m_{K^*}) A^{BK^*}_{1}(m^2_{\phi})} &\approx  &
 \frac{|V^{K^*_2\phi}_{1}(m^2_{B})|}{k(m^2_{\phi})}\,, \non\\
 \frac{(m_B+m_{K^*}) |V^{K^*\phi}_{2}(m^2_{B})|}{ A^{BK^*}_{2}(m^2_{\phi})} &\approx  &
 \frac{|V^{K^*_2\phi}_{2}(m^2_{B})|}{b_{+}(m^2_{\phi})}\,,\non \\
 \frac{(m_B+m_{K^*}) |A^{K^*\phi}(m^2_{B})|}{ V^{BK^*}(m^2_{\phi})} &\approx  &
 \frac{|A^{K^*_2\phi}(m^2_{B})|}{h(m^2_{\phi})}\,, \label{eq:ansatz}
 \ed
where we regard that the ratios on both sides have removed the
detailed characters of the different decay modes. From
Eq.~(\ref{eq:anni1}), we  know that
 $Z^{(B,P_1P_2)}_{1}=-i f_{B} (m^2_1 -m^2_2)F^{P_1
P_2}_{0}(m^2_{B})$. The explicit suppression factor
$m^2_{1}-m^2_{2}$ reflects the effects similar
to the chirality flipping on $\pi^{-} \to \mu \bar \nu_{\nu}$. Since
the dependence should be universal, although the definitions of the
time-like form factors for $\la K^{(*)} \phi| \bar{s} (V-A)_{\mu}|0
\ra$ and $\la K^{*}_n \phi| \bar{s} (V-A)_{\mu}|0 \ra$ do not
display it explicitly, for our further numerical analysis,
we  reparametrize the form factors to
have such  behavior. In addition,
to remove the ambiguity in the sign $m^2_{1}-m^2_{2}$, we set
$m_1=m_{\phi}$ and $m_{2}=(m_{K^{(*)}},\, m_{K^*_n})$. Obviously,
due to $m_{K^*_{0(2)}}> m_{\phi}>m_{K^{(*)}}$, the time-like form factors
for $K^*_{0(2)}\phi$ and $K^{(*)}\phi$, are opposite in sign. Based
on the above consideration, we could re-express
Eq.~(\ref{eq:ansatz}) by
  \be
 \frac{\left(m^2_{\phi}-m^2_{K} \right) \tilde{A}^{K
\phi}}{F^{BK}_{1}(m^2_{\phi})}&\approx &
-\frac{\left(m^2_{\phi}-m^2_{K^*_{0}} \right)\tilde{A}^{K^*_0
\phi}}{F^{BK^*_0}_{1}(m^2_{\phi})}\,,  \label{eq:reansatz_K}\\
 \frac{m^2_{B} \left(m^2_{\phi}-m^2_{K} \right)
 \tilde{V}^{K^*\phi}_{1}}{(m_B+m_{K^*}) A^{BK^*}_{1}(m^2_{\phi})} &\approx  &
 -m_{B}\frac{\left(m^2_{\phi}-m^2_{K^*_{0}} \right)\tilde{V}^{K^*_2\phi}_{1}}{k(m^2_{\phi})}\,, \non\\
 (m_B+m_{K^*})\frac{ \left(m^2_{\phi}-m^2_{K} \right)\tilde{V}^{K^*\phi}_{2}}{ A^{BK^*}_{2}(m^2_{\phi})} &\approx  &
 -m_{B}
 \frac{\left(m^2_{\phi}-m^2_{K^*_{0}} \right)\tilde{V}^{K^*_2\phi}_{2}}{b_{+}(m^2_{\phi})}\,,\non \\
 (m_B+m_{K^*})\frac{ \left(m^2_{\phi}-m^2_{K} \right)\tilde{A}^{K^*\phi}}{ V^{BK^*}(m^2_{\phi})} &\approx  &
-m_{B} \frac{\left(m^2_{\phi}-m^2_{K^*_{0}}
\right)\tilde{A}^{K^*_2\phi}}{h(m^2_{\phi})}\,,
\label{eq:reansatz_Ks}
 \ed
where the quantities with tildes at the top denote the new
 unknown parameters.
 However, in our above ansatz, the independent unknown
parameters are only $\tilde{A}^{K^{(*)} \phi}$ and  $\tilde{V}^{K^*
\phi}_{1,2}$.

 Before performing the detailed numerical calculations,
  it is
worth to show the behaviors of the polarized amplitudes and their
relationships with the annihilation effects in more concise
expressions. 
In terms of the large energy
effective theory (LEET),
we may simplify the $B\to K^*$ form factors to
be \cite{LEET}
 \be
  V(q^2)&=&(1+r_{K^*}){ \xi_{\perp}}\,, \ \ \
A_{1}(q^2)=\frac{2E}{m_{B}+m_{K^*}}\xi_{\perp}\,, \non \\
A_{2}(q^2)&=&(1+r_{K^*})\left\{\xi_{\perp}-
\frac{m_{K^*}}{E}{ \xi_{\parallel}}\right\}\,, \non \\
A_{0}(q^2)&=&(1-\frac{m^2_{K^*}}{m_{B}E})\xi_{\parallel}+r_{K^*}\xi_{\perp}\,,
 \ed
where $E$ is the energy of the $K^*$ meson, $\xi_{\parallel(\perp)}$
denotes the parallel (perpendicular) transverse form factor for
$B\to K^*$ and $r_{K^*}=m_{K^*}/m_B$. From
Eqs.~(\ref{eq:pol-amp}) and (\ref{eq:ampf_Ksphi}), the polarized
amplitudes for $B_{d}\to K^{*0} \phi$ are given by
 \be
   A_{L}&=& C_{A} \left[\xi_{\parallel} + \frac{m_{B}}{2m_{K^*}}R_a
   \left(2V^{K^*\phi}_1(m^2_{B})-V^{K^*\phi}_{2}(m^2_{B}) \right)
   \right]\,, \non \\
   A_{\parallel}&=& \sqrt{2} C_A r_{\phi} \left[\xi_{\parallel}
   + 2 R_a V^{K^*\phi}_{1}(m^2_{B}) \right]\,, \non \\
   A_{\perp}&=& \sqrt{2} C_A r_{\phi} \left[\xi_{\perp} + 2 R_a A^{K^*\phi}(m^2_{B})
   \right] \label{eq:leet_pol_amp}
 \ed
with $C_A=G_{F} V_{tb} V^{*}_{ts} \tilde{a}^{(s)} m^2_{B}
f_{\phi}/\sqrt{2}$ and $r_{\phi}=m_{\phi}/m_{B}$. From the above
equations, it is clear that $A_{L}:A_{\parallel}:A_{\perp}\approx 1:
r_\phi: r_{\phi}$. Hence, the transverse polarizations have a power
suppression in $r^2_{\phi}$. According to our previous analysis, we
 conclude that the annihilated effects and $A_{\parallel(\perp)}$ are
 in the same power. In addition, 
 by Eq.~(\ref{eq:leet_pol_amp}) we can obtain further
information on
$2V_{1}^{K^*\phi}(m^2_{B})-V_{2}^{K^*\phi}(m^2_{B})$. That is, besides the
properties displayed in Eq.~(\ref{eq:reansatz_Ks}), we expect
that $2V_{1}^{K^*\phi}(m^2_{B})-V_{2}^{K^*\phi}(m^2_{B}) =
C_{\chi}(m^2_{K^*}/m^2_{B}) (m^{2}_{\phi}-m^2_{K^*})/m^2_B $ where
$C_{\chi}$ is an unknown parameter,
which leads to
 \be
  |A_{L}|^2 \propto |\xi_{\parallel}|^2 \left( 1+ C_{\chi} R_a \frac{m_{K^*}}{m_{B}}
  \frac{m^2_{\phi}-m^2_{K^*}}{m^2_{B} \xi_{\parallel}} \right)\,.
 \ed
Clearly, the annihilation effects on $|A_{L}|^2$ are associated with
$m_{K}/m_{B}$, while those on $|A_{\parallel(\perp)}|^2$ are
$m^2_{\phi}/m^2_{B}$. By this analysis, we speculate that if
the annihilation topology in $A_{L}$ is destructive interference with
the emission one, the puzzle on the small value of 
$R_{L}(B\to K^* \phi)$ could be
solved. Similar conclusions 
could be applied to the
decays of $B\to K^*_2(1430) \phi$. 
On the contrary, if the interference
in $B\to K^*_2 \phi$ is constructive, we will get a large value of
$R_{L}(B\to K^*_{2} \phi)$. 

\begin{figure}[htbp]
\includegraphics*[width=4.in]{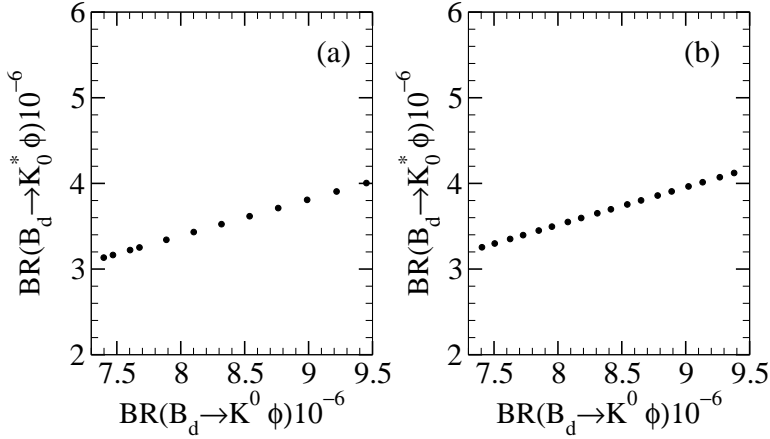}
\caption{ Correlations
between $BR(B_{d}\to K^0
\phi)$ and $BR(B_{d}\to  K^{*0}_{0}(1430) \phi)$ at $\mu=2.5$ GeV with
(a) $N^{\rm eff}_{c}=2$ and (b) $N^{\rm eff}_{c}=3$.}
 \label{fig:bpv}
\end{figure}

We now proceed 
our numerical analysis. 
Since the characters of  $B_d\to (K^0, K^{*0}_{0}(1430)) \phi$
and $B_d\to (K^{*0}, K^{*0}_{2}(1430)) \phi$ are quite different, we
discuss them separately. First, we analyze the decays of $B_d\to (K^0,
K^{*0}_{0}(1430)) \phi$. For numerical estimations, besides
the input values displayed in Tables~\ref{tab:decay_con} and
\ref{tab:ff}, to fix the unknown parameter $\tilde{A}^{K \phi}$, we
take the world average value with $1\sigma$ error for $BR(B_d\to
K^0 \phi)$, i.e.
 \be
7.3 \leq BR(B_d\to K^0 \phi)10^{6}\leq 9.5\,.
  \ed
By using Eqs.~(\ref{eq:ampf_Kphi}) and (\ref{eq:reansatz_K}), the
correlation in BRs between $B_d\to K^{*0}_{0}(1430)\phi$ and $B_d\to
K^{0}\phi$ decays is presented in Fig.~\ref{fig:bpv}, where
(a) and (b) denote the cases of $N^{\rm eff}_{c}=2$ and $3$,
respectively. From the figure, we see clearly that the results are
similar in both cases and they are consistent with the observations
by BABAR. In addition, due to the signs of the annihilation
contributions being opposite in $K^0 \phi$ and $K^{*0}_{0}(1430)
\phi$ modes, from the figure we also see that the BRs of $B_d \to
K^0 \phi$ and  $B_d \to K^{*0}_0(1430) \phi$ increase
simultaneously.

\begin{figure}[htbp]
\includegraphics*[width=5.in]{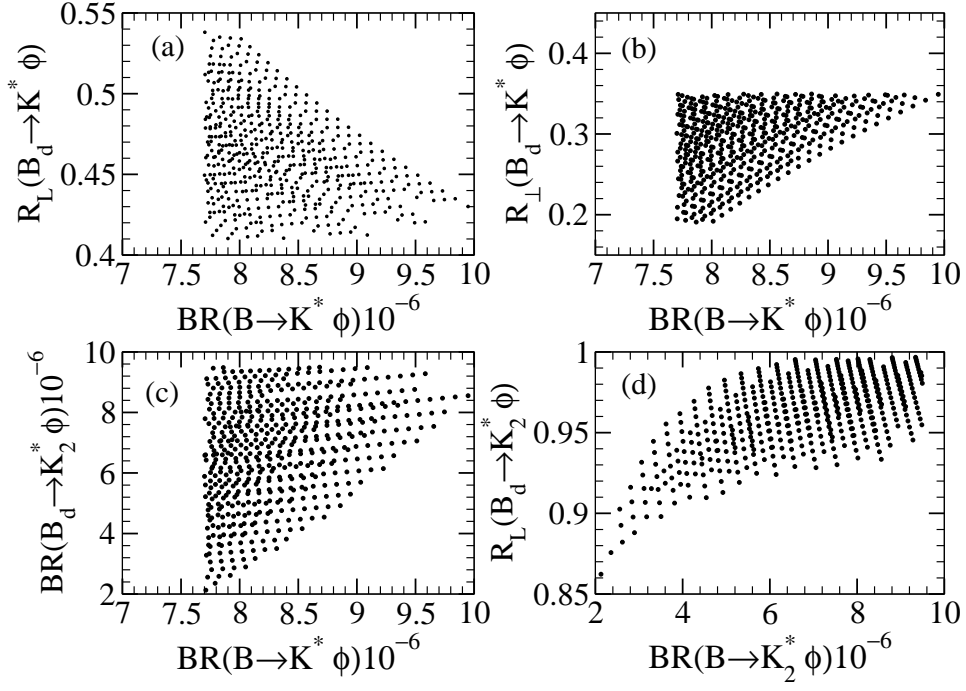}
\caption{
Correlations between (a)[(b),\,(c)] $BR(B_d\to
K^{*0}\phi)$ and $R_{L}(B_d\to K^{*0}\phi)[R_{\perp}(B_d\to
K^{*0}\phi),\, BR(B_d\to K^{*0}_{2}(1430)\phi)]$ and (d)
$R_L$ and  $BR(B_d\to K^{*0}_{2}(1430)\phi))$ in the LFQM.}
 \label{fig:btv_LF}
\end{figure}

\begin{figure}[htbp]
\includegraphics*[width=5.in]{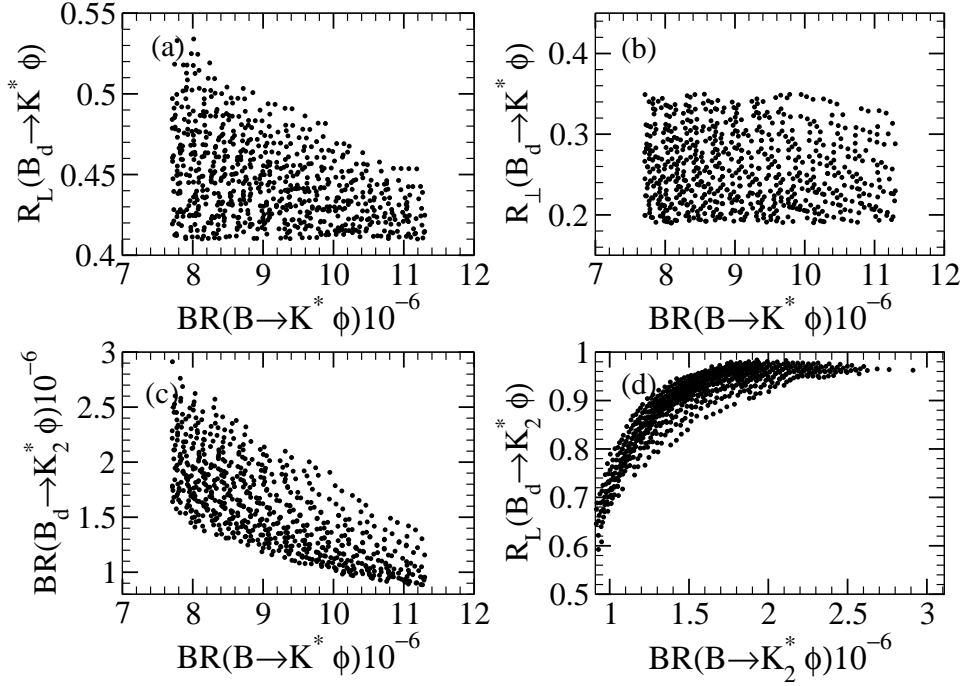}
\caption{Legend is the same as Fig.~\ref{fig:btv_LF} but
in the ISGW2.}
 \label{fig:btv_ISGW2}
\end{figure}

Next, we consider the decays of $B_{d}\to (K^{*0},\, K^{*0}_{2}(1430) )
\phi$. Similar to
$B_{d} \to  K^{0}\phi$,
to determine
$\tilde{A}^{K^{*} \phi}$ and
$\tilde{V}^{K^* \phi}_{1,2}$, we use the world average values with
$2\sigma$ errors for BR and PFs and $1\sigma$ errors for relative
angles as
 \be
  &&7.7 \leq BR(B_d\to  K^{*0} \phi)10^{6}\leq 11.3  \,, \non
 \\
 && 0.45\leq R_L(B_d\to  K^{*0} \phi)\leq 0.53\,, \ \ \ 0.21\leq R_{\perp}(B_d\to  K^{*0}
 \phi)\leq 0.35\,, \non\\
 && 2.25\leq \phi_{\parallel}( {\rm
 rad})\leq 2.59\,,\ \ \ 2.35\leq \phi_{\perp}({\rm
 rad})\leq 2.69 \,,
 \label{Constraints}
  \ed
where the angles are defined by
$\phi_{\parallel(\perp)}=Arg(A_{\parallel(\perp)}/A_{L})$. In terms
of the constraints in Eq. (\ref{Constraints}) and
Eqs.~(\ref{eq:hamp}), (\ref{eq:pol}), (\ref{eq:ampf_Ksphi}),
(\ref{eq:ampf_K2phi}) and Eq.~(\ref{eq:reansatz_Ks}), we present the
results with the form factors in the LFQM (ISGW2) and $N^{\rm
eff}_{c}=3$ in Fig.~\ref{fig:btv_LF} (\ref{fig:btv_ISGW2}).  In both
figures, the plots (a) [(b)] denote the correlations between
$BR(B_{d}\to K^{*0} \phi)$ and $R_{L}(B_{d}\to  K^{*0}
\phi)[R_{\perp}(B_{d}\to  K^{*0} \phi)]$. We note our results, which
are consistent with data, indicate that the annihilated effects are
important for the PFs. The plots (c)[(d)] display the correlations
between $BR(B_{d}\to K^{*0}_2(1430) \phi)$ and $BR(B_{d}\to K^{*0}
\phi)[R_{L}(B_{d}\to K^{*0}_2(1430) \phi)]$.
 From
 Figs.~\ref{fig:btv_LF}c and \ref{fig:btv_ISGW2}c,
we see that  $BR(B_d\to K^{*0}_{2}(1430)\phi)$ by the LFQM is much
larger than that in the ISGW2. Moreover,
the former is more favorable to the
BABAR's observation.
 From Figs.~\ref{fig:btv_LF}d and \ref{fig:btv_ISGW2}d,
 although $R_{L}$ in the ISGW2 could be
 lower,
 with the BR observed by BABAR, both
 QCD approaches  predict a large longitudinal polarization in
 $B_{d} \to K^{*0}_{2}(1430)\phi$. According to our analysis,
 we conclude that  $R_L(B \to K^{*}_2(1430) \phi)$ is
 $O(1)$ with or without
  including the annihilation contributions.
Finally, we note that the results
of $N^{\rm eff}_{c}=2$ are similar to those of
$N^{\rm eff}_{c}=3$.

\section{Conclusions}

We have studied the BRs for $B_d \to (K^{(*)0},\,
K^{*0}_{n}(1430)) \phi$ decays and the PFs for $B_d \to (K^{*0},\,
K^{*0}_{2}(1430)) \phi$ with and without the
 annihilation contributions.
Since the QCD calculations on the
 time-like form factors are not as good as
the transition form factors, we regard them as  parameters fixed by
the experimental measurements in $B_d\to K^{(*)0} \phi$. In terms of
the ansatz by correlating the time-like form factors of $\la
K^*_n(1430) \phi| (V-A)_{\mu}|0\ra$ to those of $\la K^{(*)} \phi|
(V-A)_{\mu}|0\ra$, we find that $BR(B_d\to K^{*0}_{0}(1430) \phi)$
is $(3.86 \pm 0.73)\times 10^{-6}$ and because of the annihilation
contributions to $B\to K^*_2\phi$ and $B\to K^*\phi$ being opposite
in sign, the longitudinal polarization of $B_d\to K^{*0}_2(1430)
\phi$ is always $O(1)$ unlike those in $B\to K^* \phi$ decays. Due
to the large differences in the transition form factors of $B\to
K^*_2$ between the LFQM and ISGW2, we have shown that the former
gives a broad allowed $BR(B_d\to K^{*0}_{2}(1430)\phi)$ and the
latter limits it to be $(1.69\pm 0.81)\times 10^{-6}$. In terms of
the recent BABAR's observations on BRs and PFs of $B_d\to
K^{*0}_{2}(1430)\phi $,
the results based on the LFQM are more favorable.\\

{\bf Acknowledgments}\\

This work is supported in part by the National Science Council of
R.O.C. under Grant
 \#s:NSC-95-2112-M-006-013-MY2
 and NSC-95-2112-M-007-059-MY3.


\begin{thebibliography}{99}

\bibitem{belle1}BELLE Collaboration, J. Zhang {\it et al.} , Phys. Rev. Lett. {\bf 91},
221801 (2003); K. Abe {\it et al.}, arXiv:hep-ex/0507039.

\bibitem{belle_crho} BELLE Collaboration, A. Somov {\it et al.}, Phys. Rev. Lett. {\bf 96}, 171801 (2006).


\bibitem{babar1-1}BABAR Collaboration, B. Aubert {\it et al.}, Phys. Rev. Lett.
{\bf 91}, 171802 (2003); A. Gritsan, arXiv:hep-ex/0409059.

\bibitem{babar1-2}BABAR Collaboration, B. Aubert {\it et al.}, Phys. Rev. D{\bf 69}, 031102 (2004);
Phys. Rev. Lett. {\bf 93}, 231801 (2004); B. Aubert {\it et al.},
Phys. Rev. D{\bf 71}, 031103 (2005).


\bibitem{belle2} BELLE Collaboration, K. Abe {\it et al.}, Phys. Lett. B{\bf 538}, 11 (2002);
arXiv:hep-ex/0408104.


\bibitem{babar2} BABAR Collaboration, B. Aubert {\it et al.},
Phys. Rev. Lett. {\bf 87}, 241801 (2001).

\bibitem{belle3}BELLE Collaboration, K. F. Chen, {\it et al.}, Phys. Rev. Lett. {\bf 94}, 221804 (2005).

\bibitem{babar3}BABAR Collaboration,  B. Aubert {\it et al.},
arXive:hep-ex/0303020; B. Aubert {\it et al.}, Phys. Rev. Lett. {\bf
93}, 231804 (2004).


\bibitem{Kagan}A. Kagan, Phys. Lett. B{\bf 601}, 151 (2004).

\bibitem{QCD}W.S. Hou and M. Nagashima,
hep-ph/0408007; P. Colangelo, F. De Fazio and T.N. Pham, Phys. Lett.
B{\bf 597}, 291 (2004); M. Ladisa {\it et al.},  Phys. Rev. D{\bf
70}, 114025 (2004); H.Y. Cheng, C.K. Chua and A. Soni, Phys. Rev.
D{\bf 71}, 014030 (2005); H.N. Li, Phys. Lett. B{\bf 622}, 63
(2005).

\bibitem{Chen}C.H. Chen, arXive:hep-ph/0601019.

\bibitem{Beneke}M. Beneke, J. Rohrer and D. Yang,
arXive:hep-ph/0612290.

\bibitem{newphys} A. Kagan, hep-ph/0407076; E. Alvarez {\it et al}, Phys. Rev. D{\bf 70},
115014 (2004); Y.D. Yang, R.M. Wang and G.R. Lu, Phys. Rev. D{\bf
72}, 015009 (2005); A.K. Giri and R. Mohanta, hep-ph/0412107; P.K.
Das and K.C. Yang, Phys. Rev. D{\bf 71}, 094002 (2005); C.S. Kim and
Y.D. Yang, arXiv:hep-ph/0412364, C.S. Hung {\it et al.}, Phys. Rev.
D{\bf 73}, 034026 (2006); S. Nandi and A. Kundu, J. Phys. G{\bf 32},
835 (2006) S. Baek {\it et al.}, Phys. Rev. D{\bf 72}, 094008
(2005); A.R. Williamson and J. Zupan, Phys. Rev. D{\bf 74}, 014003
(2006); Erratum {\it ibid} D{\bf 74}, 03901 (2006); Qin Chang, X.Q.
Li and Y.D. Yang, arXive:hep-ph/0610280.

\bibitem{CG_PRD71}C.H. Chen and C.Q. Geng, Phys. Rev. D{\bf 71}, 115004
(2005).

\bibitem{babar4}BABAR Collaboration,  B. Aubert {\it et al.},
arXive:hep-ex/0610073.

\bibitem{BBL} G. Buchalla, A.J. Buras and M.E. Lautenbacher, Rev.
Mod. Phys. {\bf 68}, 1125 (1996).

\bibitem{CKM} N. Cabibbo, Phys. Rev. Lett. {\bf 10}, 531 (1963); M. Kobayashi and T. Maskawa, Prog. Theor. Phys.
{\bf 49}, 652 (1973).

\bibitem{Bjorken} J. D. Bjorken, Nucl. Phys. Proc. Suppl. {\bf 11}, 325 (1989).

\bibitem{WF_QCD}P. Ball and R. Zwicky, Phys. Rev. D{\bf 71}, 014015
(2005); {\it ibid}, {\bf 71}, 014029 (2005); Phys. Lett. B{\bf 633},
289 (2006); JHEP {\bf 0602}, 034 (2006); JHEP {\bf 0605}, 004
(2006).

\bibitem{GFA1} A. Ali, G. Kramer and C.D. Lu,
Phys. Rev. D{\bf 58}, 094009 (1998).

\bibitem{GFA2} Y.H. Chen {\it et al.}, Phys. Rev D{\bf 60},
094014 (1999).

\bibitem{BSW} M. Bauer, B. Stech and M. Wirbel, Z. Phys. C{\bf 34},
103 (1987).

\bibitem{LF1} H.Y. Cheng, C.K. Chua and C.W. Hwang, Phys. Rev. D{\bf 69}, 074025
(2004).


\bibitem{EDS}L.N. Epele, D. Gomez Dumm and  A. Szynkman, Eur.Phys. J. C{\bf 29}, 207
(2003).

\bibitem{BDDN} E.R. Berger, A. Donnachie, H.G. Dosch and O. Nachtmann,
Eur.Phys. J. C{\bf 14}, 673 (2000).

\bibitem{BVV}G. Kramer and W.F. Palmer, Phys. Rev. D{\bf 45}, 193
(1992).

\bibitem{ISGW2} N. Isgur, D. Scora, B. Grinstein and M.B. Wise, Phys. Rev. D{\bf 39}, 799 (1989);
D. Scora and N. Isgur, Phys. Rev. D{\bf 52}, 2783
(1995).

\bibitem{KLO_PRD67}C.S. Kim, J.P. Lee and S. Oh, Phys. Rev. D{\bf
67}, 014002 (2003).


\bibitem{CGHW}C.H. Chen, C.Q. Geng, Y.K. Hsiao and Z.T. Wei, Phys. Rev. D{\bf 72}, 054011
(2005).

\bibitem{PQCD1} G.P. Lepage and S.J. Brodsky, Phys. Lett. B{\bf 87}, 359 (1979);
Phys. Rev. D{\bf 22}, 2157 (1980); H.N. Li and G. Sterman, Nucl.
Phys. B{\bf 381}, 129 (1992); G. Sterman, Phys. Lett. B{\bf 179},
281 (1986); Nucl. Phys. B{\bf 281}, 310 (1987); S. Catani and L.
Trentadue, Nucl. Phys. B{\bf 327}, 323 (1989); Nucl. Phys. B{\bf
353}, 183 (1991); Y.Y. Keum, H.N. Li and A.I. Sanda, Phys. Rev.
D{\bf 63}, 054008 (2001); H.N. Li, Phys. Rev. D{\bf 64}, 014019
(2001); H.N. Li, Phys. Rev. D{\bf 66}, 094010 (2002).

\bibitem{PQCD2} T.W. Yeh and H.N. Li, Phys. Rev. D{\bf 56}, 1615
(1997).

\bibitem{CKLPRD66} C.H. Chen, Y.Y. Keum and H.N. Li, Phys. Rev.
D{\bf 66}, 054013 (2002).

\bibitem{SCET}C.M. Arnesen, I.Z. Rothstein and I.W. Stewart,
arXive:hep-ph/0611356.


\bibitem{LEET} M.J. Dugan and B. Grinstein, Phys. Lett. B{\bf 255}, 583 (1991); J. Charles {\it
et al.}, Phys. Rev. D{\bf 60}, 014001 (1999).


\end{thebibliography}
\end{document}